\RequirePackage[T1]{fontenc}
\documentclass[conference]{IEEEtran}
\usepackage{cite}
\usepackage{amsmath,amssymb}
\usepackage{graphicx}
\usepackage{booktabs}
\usepackage{url}
\usepackage[hidelinks]{hyperref}
\hypersetup{pdfauthor={},pdfsubject={},pdfkeywords={},pdfcreator={},pdfproducer={}}
\newcommand{\doilink}[1]{\href{https://doi.org/#1}{\nolinkurl{#1}}}
\newcommand{\tstar}{t^{*}}

\begin{document}
\raggedbottom
\title{Conformal Risk Control for Sentinel-1 Flood Maps\newline
under Event Shift and Local Recalibration}

\author{
\IEEEauthorblockN{%
Izzath Nisfer, 
Rahim Iqbal, 
Shazan Shaheed, 
Asfi Ahamed, 
Muhammadu Ilham,\\
Dileesha Manamperi, 
Thimira Sahan, 
Sandareka Wickramanayake
}
\IEEEauthorblockA{Dept.\ of Computer Science \& Engineering, University of Moratuwa, Sri Lanka.\\
\texttt{\{rahimi.23, asfia.23, izzathn.23, shazans.23, muhammadui.23,}\\
\texttt{dileesha.26, thimira.22, sandarekaw\}@cse.mrt.ac.lk}
}
}

\maketitle

\begin{abstract}
Conformal Risk Control (CRC) selects a threshold to limit missed labeled positives under exchangeability, but low omission risk can come at the cost of an excessively large flood map. We examine this trade-off with a frozen bi-temporal Sentinel-1 predictor, 403 expert-labeled source groups, and ten evaluable leave-one-event-out folds. At a target risk of 0.10, source-calibrated thresholds produce mean group losses of 0.1726 in Somalia and 0.1985 in Nigeria (four groups). The other eight folds select zero thresholds, labeling every valid pixel as flooded; their excess predicted area is 3.15--20.55 times the labeled positive area. Local threshold recalibration is feasible in 559 of 570 scheduled splits. The equal-event frequency of meeting the observed risk target with a positive threshold is 0.95\% at ten labeled groups and 8.00\% at twenty, using different predefined event pools. At fifty groups, exploratory recalibration meets both conditions in 24/30 Paraguay splits and none in India or the United States. Labels include permanent water, so the loss does not isolate newly flooded land. The results show why omission risk should be reported alongside mapped area, calibration feasibility, and event-specific outcomes.

\end{abstract}

\begin{IEEEkeywords}
Conformal risk control, distribution shift, flood mapping, recalibration,
Sentinel-1.
\end{IEEEkeywords}

\section{Introduction}
Synthetic Aperture Radar (SAR) observes the surface through clouds and without daylight, making it valuable during floods. Vegetation, buildings, surface roughness, and acquisition geometry nevertheless complicate flood delineation \cite{amitrano2024_sarreview}. Good segmentation on one event therefore need not translate into reliable maps elsewhere.

\begin{figure*}[!t]
\centering
\begin{tabular}{ccc}
\small (a) Expert reference & \small (b) Errors at $t=0.5$ & \small (c) Errors at $t=1/512$ \\[2pt]
\includegraphics[width=0.25\textwidth,trim=26.87778bp 28.79762bp 881.68711bp 59.03512bp,clip]{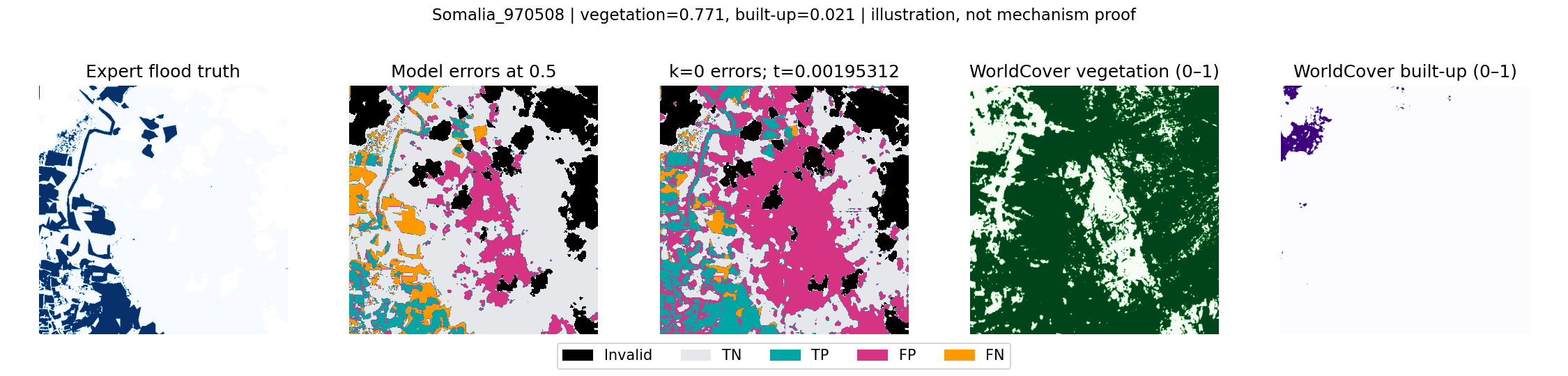} &
\includegraphics[width=0.25\textwidth,trim=240.46012bp 28.79762bp 668.10477bp 59.03512bp,clip]{figures/c6_a2_Somalia_970508_overlay.png} &
\includegraphics[width=0.25\textwidth,trim=454.52243bp 28.79762bp 454.04247bp 59.03512bp,clip]{figures/c6_a2_Somalia_970508_overlay.png}
\end{tabular}\\[3pt]
\includegraphics[width=0.48\textwidth,trim=381.56846bp 2.39980bp 381.56846bp 235.18056bp,clip]{figures/c6_a2_Somalia_970508_overlay.png}
\caption{Somalia\_970508 (38,112 positive / 217,119 valid pixels). Blue: reference positives; TP/TN: true positives/negatives; FP/FN: false positives/negatives. This deterministic post-analysis illustration is not representative.}
\label{fig:somalia}
\end{figure*}

Conformal Risk Control (CRC) selects an operating parameter for a bounded monotonic loss \cite{crc}. Lowering a flood probability threshold reduces missed positives while expanding the predicted area. Under exchangeability of calibration and future groups, CRC controls expected loss. That assumption is uncertain when transferring between events. Moreover, predicting every valid pixel as flooded eliminates omission without providing useful boundaries. Fig.~\ref{fig:somalia} illustrates this trade-off for one held-out group.

We evaluate source threshold transfer and local recalibration using a fixed bi-temporal predictor and expert labels from Modified Sen1Floods11 \cite{modified_sen1floods11_2023}. Local labels change only the threshold; network weights remain fixed. Because the positive class includes permanent water, ``missed-flood risk'' here means missed retained positive labels, rather than omission of newly flooded land alone.

The study addresses three questions:
\begin{itemize}
\item Does source calibration meet the observed risk target on a held-out event, and how much area does the resulting map label positive?
\item How often do fixed local labeling budgets recover a positive threshold that also meets the observed target?
\item How does the bi-temporal predictor compare with a matched post-event-only model in segmentation quality and computational cost?
\end{itemize}
Our contribution is an empirical assessment of transfer, conservatism, and recalibration costs using existing CRC methods. Keeping the network fixed separates threshold adjustment from changes in the learned representation.

\section{Related Work}
Sen1Floods11 supplies georeferenced Sentinel-1 observations and expert surface-water labels \cite{sen1floods11_2020}; its modified release adds before-and-after pairs \cite{modified_sen1floods11_2023}. We use these pairs in a compact U-Net-style predictor \cite{unet2015} to study threshold transfer.

Conformal methods already support flood mapping. Konidakis \emph{et al.}~\cite{konidakis2025} evaluate inductive and cross-validation conformal prediction for bi-temporal Sentinel-2 segmentation. Li \emph{et al.}~\cite{li2026_sarcrc} combine density-aware SAR segmentation, out-of-distribution abstention, and set-valued predictions with calibration-conditional risk control, reporting risk exceedance in complex mismatched cases. We instead evaluate a single threshold using mean group omission loss; the different operating rules and evaluation units preclude direct performance comparisons.

Weighted conformal prediction addresses covariate shift under an unchanged conditional response distribution and a known covariate likelihood ratio \cite{weightedcp2019}. Non-exchangeable CRC relates risk deterioration to distributional differences and permits relevance weighting \cite{nonxcrc2024}. These results do not certify our unweighted procedure under arbitrary event shift.

Local repair is also established outside flood mapping. L\"ofstr\"om~\cite{repairbudget2026} studies regression interventions under empirical coverage and interval-width constraints. Adhikary \emph{et al.}~\cite{adhikary2026_boundaware} examine transferred recall thresholds, vacuous solutions, and local recertification in medical segmentation. We assess these practical issues across SAR flood events, retaining unsuccessful recalibration attempts and reporting mapped area alongside omission risk.

\section{Data and Experimental Design}

\subsection{Data Selection and Evaluation Unit}
We use the expert-labeled portion of Modified Sen1Floods11 v1~\cite{modified_sen1floods11_2023}, which adds pre-event imagery to the original Sen1Floods11 dataset~\cite{sen1floods11_2020}. Of 446 expert identities, 28 lack pre-event images and 15 of the remaining 418 triplets have truncated pre-event geometry, leaving 403 eligible source groups. Each $512\times512$ group is processed as four $256\times256$ quadrants that retain the original identity and allocation role. The 1,612 quadrants are therefore not independent calibration samples.

Table~\ref{tab:events} summarizes the held-out support. Eleven fold definitions yield ten evaluable events: Bolivia has no eligible groups, and Nigeria has only four. Five Ghana groups have no jointly valid pixels, leaving 398 usable groups for CRC. Nominal label budgets can therefore exceed usable calibration counts.

Each acquisition provides VV backscatter, with vertical transmission and vertical reception, and VH backscatter, with vertical transmission and horizontal reception. Both channels follow the fixed pipeline's decibel (dB) convention. Valid pixels require finite, unmasked values in both channels at both dates and an expert label of 0 or 1. Invalid inputs are zero-filled after normalization and excluded from losses. Triplets must match in dimensions, coordinate reference system, and affine grid; these checks do not establish sub-pixel registration accuracy or identical viewing geometry.

\subsection{Event Separation and Predictor Training}
Each leave-one-event-out fold excludes the target event from training and model selection. The fixed allocation manifest reserves approximately 20\% of expert groups within each source event for CRC calibration. The remaining outer training pool is split for fitting and inner validation. Normalization uses the outer pool, including inner validation, while excluding CRC calibration and target groups. CRC labels do not select checkpoints. Sen1Floods11 also provides weakly labeled chips~\cite{sen1floods11_2020}, but these provide neither calibration nor evaluation truth. Training applies matched flips and quarter-turn rotations to both dates, labels, and masks.

This allocation separates three uses of labels: fitting network weights, choosing a checkpoint on inner validation, and selecting the CRC threshold. The target labels enter only the later recalibration analysis. All four quadrants of a source group retain the same role, preventing fragments of one image identity from appearing in different pools. The calibration sample size therefore counts usable source identities, rather than tiles or pixels.

The temporal core shares an encoder across dates, with feature widths 16, 32, 64, 128, and 256 and one decoder. Given pre-event features $P_i$ and post-event features $S_i$, the fused skip connection is
\begin{equation}
F_i=S_i+s_{i0}(S_i-P_i)+s_{i1}|S_i-P_i|
    +s_{i2}g_i\odot Q_i([P_i,S_i]),
\label{eq:fusion}
\end{equation}
where $Q_i$ is a depthwise-convolution, ReLU, and pointwise-convolution projection, and $g_i=\sigma(W_i|S_i-P_i|+b_i)$ is a learned gate. Here, $\sigma$ is the sigmoid function and $\odot$ denotes elementwise multiplication. The learned scalars $s_{ij}$ start at $1/3$ and are not softmax weights. The core uses no permanent-water prior, consistency loss, batch normalization, or Transformer blocks. The matched post-event-only baseline uses the same encoder-decoder family and base width 16.

Both models use seed 42, batch size 8, 25 epochs, FP32, and AdamW~\cite{adamw2017} with learning rate $10^{-3}$, weight decay $10^{-4}$, and cosine annealing without restarts. The objective sums equally weighted focal~\cite{focal2017} and smoothed soft Dice losses over valid batch pixels. Focal loss uses $\gamma=2$ and class weights 0.5; Dice loss is $1-(2\sum py+1)/(\sum p+\sum y+1)$, a non-squared-denominator variant of differentiable Dice objectives~\cite{vnet2016}. Checkpoints maximize inner-validation flood Intersection over Union (IoU). Standard segmentation is evaluated at $t=0.5$ by pooling valid-pixel confusion counts within each event and averaging the resulting event IoUs equally, whereas CRC risk averages group-level missed-positive losses.

The focal term is averaged over valid pixels, while the Dice sums cover those same pixels across the batch. Each term has coefficient one. Our Dice denominator differs from the squared-denominator formulation in V-Net \cite{vnet2016}; the citation supplies background for the objective, rather than an exact implementation identity.

\begin{table*}[!t]
\centering
\caption{Event support, standard segmentation, and source-calibrated risk. Seg.: held-out groups in segmentation evaluation; CRC: held-out groups with valid pixels used for risk evaluation.}
\label{tab:events}
\small
\setlength{\tabcolsep}{6pt}
\begin{tabular}{lrrrrrrrr}
\toprule
& \multicolumn{2}{c}{Held-out groups} & \multicolumn{2}{c}{IoU at $t=0.5$}
& \multicolumn{4}{c}{Source-calibrated operating point} \\
\cmidrule(lr){2-3}\cmidrule(lr){4-5}\cmidrule(lr){6-9}
Event & Seg. & CRC & Post-only & Temporal & $\tstar$ & Risk & Excess area & IoU \\
\midrule
Ghana & 50 & 45 & 0.4097 & 0.4630 & 0 & 0.0000 & 14.35 & 0.0652 \\
India & 64 & 64 & 0.5405 & 0.5327 & 0 & 0.0000 & 7.06 & 0.1241 \\
Mekong & 30 & 30 & 0.8490 & 0.8637 & 0 & 0.0000 & 3.15 & 0.2412 \\
Nigeria & 4 & 4 & 0.5408 & 0.5505 & $1/512$ & 0.1985 & 3.62 & 0.1829 \\
Pakistan & 28 & 28 & 0.2684 & 0.2860 & 0 & 0.0000 & 12.34 & 0.0750 \\
Paraguay & 65 & 65 & 0.5427 & 0.4974 & 0 & 0.0000 & 11.25 & 0.0816 \\
Somalia & 26 & 26 & 0.2456 & 0.2161 & $1/512$ & 0.1726 & 4.71 & 0.1414 \\
Spain & 30 & 30 & 0.7285 & 0.7412 & 0 & 0.0000 & 6.60 & 0.1315 \\
Sri Lanka & 41 & 41 & 0.5675 & 0.6441 & 0 & 0.0000 & 7.42 & 0.1187 \\
USA & 65 & 65 & 0.5601 & 0.5520 & 0 & 0.0000 & 20.55 & 0.0464 \\
\bottomrule
\end{tabular}
\end{table*}

\section{Risk Control and Local Recalibration}

\subsection{Loss and Threshold Selection} 
For a group $i$, let $V_i$ be the set of its valid pixels, $y_{iu}\in\{0,1\}$ the expert label, and $p_{iu}$ the model flooding probability. The prediction uses $p_{iu}\geq t$. The missed-positive loss is

\begin{equation} L_i(t)=\frac{\sum_{u\in V_i}y_{iu}\mathbf{1}\{p_{iu}<t\}} {\sum_{u\in V_i}y_{iu}}. \label{eq:loss} \end{equation}

Valid groups without positive labels receive $L_i(t)=0$; groups without valid pixels are excluded. Empirical risk is $\widehat R_n(t)=n^{-1}\sum_i L_i(t)$. Each usable group has equal weight regardless of positive area, so this differs from pooled pixel false-negative rate and from loss conditional on a group containing positives. We combine quadrant counts before evaluating the ratio in~\eqref{eq:loss}. Label budgets and calibration sample sizes count original groups.

Consequently, missing the same fraction of positives contributes equally for a group with a small positive region and one with a large region. Averaging quadrant-level ratios would change that weighting, especially when positives are concentrated in one quadrant. Retaining zero-positive groups also makes the mean depend on their prevalence. These choices define the reported risk and must accompany any comparison with pixel-level recall.

With loss bound $B=1$, target $\alpha=0.10$, and fixed grid $G=\{j/512:j=0,\ldots,512\}$, we select

\begin{equation} \tstar=\max\left\{t\in G: \frac{n\widehat R_n(t)+1}{n+1}\leq\alpha+10^{-12}\right\}. \label{eq:crc} \end{equation}

Loss increases with $t$, so the largest admissible threshold minimizes predicted area on the grid. Setting $\lambda=1-t$ gives CRC's usual non-increasing loss parameterization \cite{crc}. The $10^{-12}$ tolerance applies only to selection; observed evaluation risk is compared with exactly 0.10.

Since $L_i(0)=0$, the rule is feasible if and only if $n\geq9$. Smaller samples are recorded as infeasible instead of using CRC's endpoint fallback. Ignoring numerical tolerance,~\eqref{eq:crc} requires
\begin{equation}
\widehat R_n(t)\leq\alpha-\frac{1-\alpha}{n}.
\label{eq:finite}
\end{equation}
For $n=10,20,50$, the allowed empirical losses are 0.010, 0.055, and 0.082. Feasibility alone does not ensure a positive threshold: nine groups can admit one if no calibration positives are missed, while larger samples can still select zero.

CRC controls expected future-group loss under exchangeability, averaging over calibration and future groups \cite{crc}. It does not guarantee the target for every event or realized evaluation set. Source transfer is therefore an empirical stress test. Local recalibration also needs exchangeability with future target groups for a theoretical interpretation; disjoint identities do not establish it, and spatial dependence is not modeled.

\subsection{Source Transfer and Target-Label Budgets}
At $k=0$, each fold selects $\tstar$ from 66--78 usable source calibration groups before target evaluation. At $k>0$, $k$ nominal target groups recalibrate only the threshold of the frozen network; the remaining target groups form the disjoint evaluation set.

The committed plan uses 30 deterministic nested-permutation repeats per event--budget cell (root seed 20260913): 19 cells and 570 scheduled splits. The primary $k=10$ pool comprises Ghana, India, Mekong, Paraguay, Spain, Sri Lanka, and USA; the primary $k=20$ pool comprises Ghana, India, Paraguay, Sri Lanka, and USA. Exploratory companions add Pakistan and Somalia at $k=10$, and Mekong and Spain at $k=20$. All $k=50$ cells (India, Paraguay, USA) are exploratory. Budget five is infeasible under the rule, and no event supports budget 100.

Tiers were fixed on the original 418 complete triplets, requiring at least 20 nominal assessment groups for primary cells and ten for exploratory cells. They were retained after the geometry audit reduced eligibility to 403, rather than reassigned from outcomes. At $k=50$, the nominal evaluation remainder is 14 groups for India and 15 each for Paraguay and USA. Masking can further reduce support.

Primary pools differ across budgets, preventing a matched interpretation of their aggregate comparison. Larger budgets also shrink evaluation sets. Repeated splits reuse groups, so frequencies summarize split sensitivity rather than independent trials or population confidence intervals.

\begin{figure*}[!t]
\centering
\includegraphics[width=\textwidth]{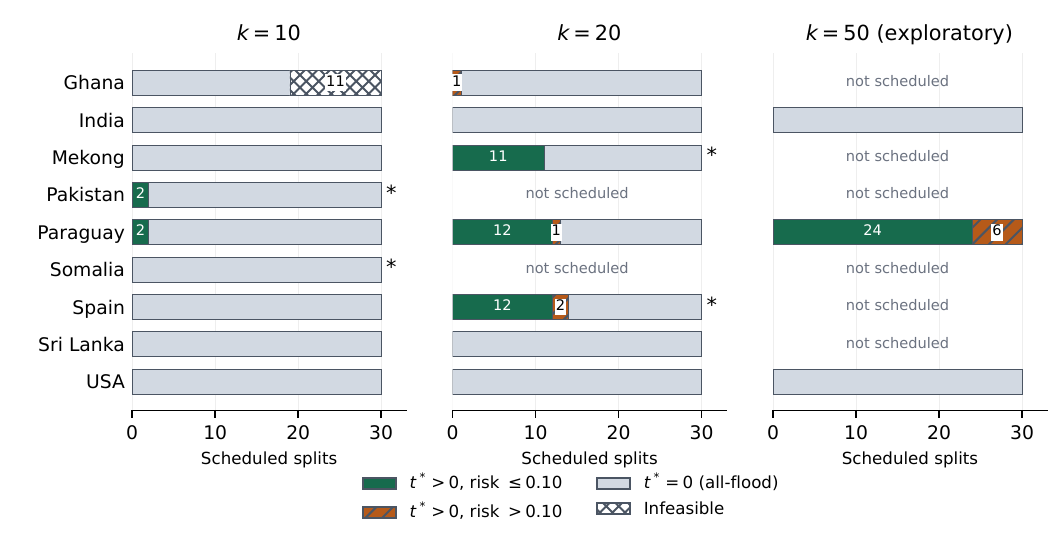}
\caption{Outcomes of all 570 scheduled splits, grouped by event and label budget. An asterisk ($*$) marks an exploratory companion cell at $k=10$ or $k=20$. All $k=50$ cells are exploratory. Counts denote splits, not independent trials.}
\label{fig:outcomes}
\end{figure*}

\subsection{Joint Risk and Informativeness Reporting}
We record whether a feasible split has mean evaluation loss $\leq0.10$, whether $\tstar>0$, and whether both hold. Infeasible splits count toward neither outcome but remain in scheduled denominators. Frequencies use all 30 scheduled splits per cell, then equal event weights within the declared pool.

A zero threshold labels every valid pixel as flooded. A positive threshold only excludes this endpoint; it does not ensure acceptable precision or map extent, and can still label every valid pixel positive if all scores exceed it. We therefore also report IoU, predicted-positive fraction where relevant, and excess area $A_{\rm pred}/A_{\rm true}-1$, counting valid predicted and labeled positive pixels. These are diagnostics, not controlled losses. For an all-flood map, IoU equals the labeled-positive fraction of valid area, explaining how zero omission can coexist with poor overlap.

\section{Results}

\subsection{Source-Calibrated Thresholds under Event Shift}
Table~\ref{tab:events} shows that Somalia and Nigeria exceed the 0.10 target at $\tstar=1/512$, with mean group losses of 0.1726 and 0.1985, respectively. Nigeria's four groups warrant descriptive interpretation. Somalia's pooled pixel recall is 0.8309, differing from $1-0.1726=0.8274$ because CRC weights groups equally.

The other eight folds select $\tstar=0$ on source data before target evaluation. Their zero omission is mechanical: every valid target pixel is labeled positive. Excess area ranges from 3.15 (Mekong) to 20.55 (USA), meaning total predicted areas of 4.15--21.55 times labeled positive area. IoU ranges from 0.0464 to 0.2412 despite higher scores at the standard threshold.

For example, the temporal predictor's Mekong IoU falls from 0.8637 at $t=0.5$ to 0.2412 at the source-calibrated threshold. In USA it falls from 0.5520 to 0.0464. These comparisons show that reasonable overlap at a conventional threshold can coexist with a highly conservative CRC operating point. They describe the cost of the selected omission objective, rather than a change in network weights.

The Somalia panel in Fig.~\ref{fig:somalia} gives a spatial view of this trade-off, showing the reference and errors at the standard and source-calibrated thresholds for the same group. It is a deterministic example selected after analysis, not an estimate of typical event performance or evidence for a physical cause of the errors.

By monotonicity, selecting zero means the source criterion fails even at $1/512$, the smallest positive grid point. This shows conservatism under the fixed predictor, loss, grid, and calibration setup; it does not show that target shift caused threshold collapse or that cutoffs between zero and $1/512$ would fail.

\subsection{Local Recalibration Is Usually Conservative}

Of 570 scheduled splits, 559 are feasible. All 11 infeasible attempts occur in Ghana at $k=10$, where masking leaves 6--8 usable calibration groups. They receive no prediction and remain in the denominator. Fig.~\ref{fig:outcomes} includes all cells, including
exploratory companions excluded from primary means.

Table~\ref{tab:repair} shows that all 199 feasible primary $k=10$ splits meet the observed target, but only two select positive thresholds. At primary $k=20$, all 150 splits are feasible; fourteen select positive thresholds and twelve meet both conditions. The equal-event joint frequencies are 0.95\% and 8.00\%, respectively, across different event pools. High target-attainment frequencies thus mainly reflect conservative thresholds.

\begin{table}[!ht]
\centering
\caption{Local recalibration outcomes by declared event pool.}
\label{tab:repair}
\small
\setlength{\tabcolsep}{3pt}
\begin{tabular}{lrrr}
\toprule
Budget $k$ & 10 & 20 & 50 \\
Analysis tier & Primary & Primary & Exploratory \\
Events & 7 & 5 & 3 \\
Scheduled / feasible & 210/199 & 150/150 & 90/90 \\
\midrule
Risk $\leq0.10$ (\%) & 94.76 & 98.67 & 93.33 \\
$\tstar>0$ (\%) & 0.95 & 9.33 & 33.33 \\
Both conditions (\%) & 0.95 & 8.00 & 26.67 \\
\bottomrule
\end{tabular}
\end{table}

Paraguay accounts for both primary joint outcomes at $k=10$ and all twelve at $k=20$. Ghana's single positive-threshold outcome at $k=20$ exceeds the risk target. Exploratory companions yield two joint outcomes in Pakistan at $k=10$, eleven in Mekong at $k=20$, and twelve in Spain at $k=20$. India and USA never select positive thresholds at any scheduled budget.

At exploratory $k=50$, Paraguay selects a positive threshold in all 30 splits and meets the observed target in 24. India and USA select zero throughout, so the 26.67\% equal-event joint frequency reflects recovery in one of three events. Paraguay's marginal split medians are threshold 0.00390625, risk 0.0773, predicted-positive fraction 0.2605, and IoU 0.2715. These medians need not describe the same map, and even this favorable case offers limited segmentation overlap.

\subsection{Predictor Quality and Computational Cost}
Using the event-level IoUs in Table~\ref{tab:events},
the equal-event mean at $t=0.5$ is 0.5253 for the
post-only model and 0.5347 for the temporal core,
a difference of $+0.0094$, with six event-level gains
and four losses. This comparison is descriptive
because each fold uses one training seed.

The temporal core uses 2,213,776 parameters versus 1,940,961 for post-only ($+14.06\%$). Matched Tesla T4 median forward-pass latency is 4.7144 versus 2.2657~ms (FP32, batch one, $256\times256$ inputs), excluding preprocessing, loading, and map assembly. The approximately twofold latency exceeds the original 25\% overhead target. These mixed accuracy and cost results do not establish that temporal fusion is necessary for the CRC findings; predictor families were not compared for transfer or recalibration.

Table~\ref{tab:controls} reports six supporting runs comparing a shared-weight Siamese difference baseline with base width 16 and a larger post-only control with width 32 and 7,759,809 parameters across three prespecified events. Post32 exceeds the core in Somalia and slightly in USA, but not Sri Lanka; Siamese16 trails it in all three. This limited comparison does not establish a general architectural ranking.

\begin{table}[!ht]
\centering
\caption{Supporting three-event controls: IoU at threshold 0.5.}
\label{tab:controls}
\small
\setlength{\tabcolsep}{5pt}
\begin{tabular}{lrrr}
\toprule
Event & Temporal core & Siamese16 & Post32 \\
\midrule
Somalia & 0.2161 & 0.1986 & 0.2818 \\
Sri Lanka & 0.6441 & 0.5660 & 0.5676 \\
USA & 0.5520 & 0.5482 & 0.5559 \\
\bottomrule
\end{tabular}
\end{table}

\section{Discussion and Limitations}
\subsection{Interpreting Risk Together with Map Extent}
Low omission risk alone is insufficient for useful flood delineation. An all-flood map meets the target while failing to distinguish wet from dry pixels. Reports should therefore give the loss, aggregation unit, usable calibration support, threshold, and mapped area together. Permanent-water labels and zero-positive groups are part of our estimand; changing either requires reevaluation.

The source and local results expose different issues. Source calibration can already select the all-flood endpoint before seeing the target, while a positive transferred threshold can exceed the target's observed risk. Local recalibration may change the operating point, but it leaves the predictor's scores unchanged. The joint outcome is therefore a useful screening measure, with IoU and area needed to assess the resulting map.

\subsection{What a Local Label Budget Can Establish}
More local labels relax the finite-sample correction, but recovery also depends on calibration positives and the evaluation remainder. The differing primary pools and dependent splits establish no universal annotation budget. Comparisons with fine-tuning, shift weighting, and other losses remain untested. Precision or excess-area constraints could make the operating objective more useful, but are not certified here.

Paraguay illustrates this distinction: fifty groups permit positive thresholds in every split, yet six of those splits exceed the observed risk target and the median IoU remains 0.2715. India and USA show no positive-threshold recovery at that budget. Thus, a larger feasible sample should not be interpreted as a promise of informative maps across events, even when the finite-sample correction is less restrictive.

\subsection{Scope of the Evidence}
Evidence is limited to one dataset, one CRC predictor family, and one seed per fold; Nigeria has four groups. Excluding incomplete or mismatched triplets narrows the population. Matching raster grids does not prove registration accuracy, and incidence-angle metadata were unavailable. Spatial dependence and changes in labeling can affect transfer; disjoint groups and observed performance do not establish future exchangeability. The fixed grid also leaves cutoffs below $1/512$ unexplored. These limitations qualify the protocol's empirical results rather than imply an impossibility result for CRC.

\section{Conclusion}
Across ten evaluable Sentinel-1 events, source-calibrated CRC exhibits two limitations: observed omission risk exceeds the target in Somalia and low-support Nigeria, while the other eight events select all-flood thresholds. Local threshold recalibration meets the observed target with a positive cutoff in only 0.95\% and 8.00\% of splits under the respective primary event pools. Exploratory $k=50$ recalibration achieves both conditions in 24/30 Paraguay splits, but not in India or USA. Thus, low omission risk does not by itself establish useful flood delineation, and additional labels do not guarantee recovery. These findings are limited to one dataset, one training seed per fold, and dependent repeated splits. Future evaluation should test whether alternative predictors and joint omission--mapped-area objectives improve transfer without concealing infeasible or degenerate outcomes.

\bibliographystyle{IEEEtran}
\bibliography{references_conference}
\end{document}